\documentclass[aps,twocolumn,preprintnumbers,nofootinbib,superscriptaddress]{revtex4-1}
\usepackage{amsmath} \usepackage{graphicx} \usepackage{amsfonts}
\usepackage{array} \usepackage{amsthm} \usepackage{bm}
\usepackage{palatino} \usepackage{mathpazo} 
\usepackage{supertabular}

\usepackage{latexsym}
\usepackage{hyperref}
\hypersetup{colorlinks=true,
	breaklinks=true,
	pdfstartview=Fit,
	linkcolor=magenta,
	citecolor=blue,
	urlcolor=blue}

\newcommand{\be}{\begin{equation}}
\newcommand{\ee}{\end{equation}}
\newcommand{\ba}{\begin{eqnarray}}
\newcommand{\ea}{\end{eqnarray}}
\newcommand{\bal}{\begin{align}}
\newcommand{\eal}{\end{align}}

\newcommand{\e}{\text{e}}
\newcommand{\dd}{\text{d}}

\newcommand{\bb}{\bibitem}

\newcommand{\bw}{\begin{widetext}}
\newcommand{\ew}{\end{widetext}}

\begin{document}
\title{The Star Universe Model -- Another FLRW Patch}

\author{Mustapha Azreg-A\"{\i}nou}
\affiliation{Ba\c{s}kent University, Engineering Faculty, Ba\u{g}l{\i}ca Campus, 06780-Ankara, T\"{u}rkiye}


\begin{abstract}
The Black Hole Universe model has advanced in the literature. It assumes that the whole mass of the Universe remains inside the Schwarzschild radius of a black hole after being subject to a spherical collapse. 
We present a graphical proof and show that within the realm of classical theory this is not possible if spacetime remains regular at all times during the evolving collapse/bounce. Other drawbacks are discussed. As an alternative, we propose the Star Universe Model. In this model no other ingredients (such as inflaton field and dark energy) other than the perfect fluid itself are considered, and a positive spatial curvature is naturally introduced along with a particle production rate function. We discuss new-old static solutions, show that the density of a semi-closed Universe of gravitational mass $M_{\text{g}}$ is bounded from above by $3c^6/(32 \pi  G^3 M_{\text{g}}^2)$, and elaborate on the birth of the Universe and calculate its inflation rate within the Star Universe Model. A smooth graceful exit from inflationary state occurs as the evolutionary solution approaches the generalized Einstein static Universe (with uniform energy density and negative pressure) and the expansion becomes linear in cosmological time with density and (negative) total pressure proportional to the square of the inverse of the scale factor. A similar behavior for density and (positive) pressure will occur if (re)collapse takes place. The analysis results in a time variable and decreasing effective cosmological constant $8\pi G\rho(\tau)/c^2$, which remains almost constant at late times, that justifies the late accelerated expansion with a rate of particle production three times the Hubble parameter.
\end{abstract}

\maketitle

\section{Introduction}
In the past, the lack of accurate data in astronomy and the lack of laboratory instruments that generate high energies to discover new more fundamental particles had led the scientific community to make bold hypotheses for the purpose of advancing science. Bekenstein was the first who made the bold conjecture concerning the entropy of a black hole (BH), which later led to the discovery of its formula by Hawking. The de Broglie wave length and Planck harmonic-oscillator energy are but other bold or ad hoc hypotheses. Today, despite the overwhelming of cosmological data and accuracy of measurements, inflaton and dark energy hypotheses are still under scrutiny; moreover, many modified or generalized theories of General Relativity (GR) do not support the existence of dark energy.

Another bold hypothesis that the Universe is inside the Schwarzschild radius of a BH was made by Raj Kumar Pathria~\cite{RKP}. This idea has always been surfacing with amendments and criticisms~\cite{HK,SK,NJP,EKI,GKPG}. Very recently, a proposal that this was actually the case has been advanced in~\cite{GKPG} and constituted the essence of the so-called Black Hole Universe (BHU). The investigation assumed that a relativistic spherical collapse of a perfect fluid with an evolving equation of state may be halted, preventing the formation of a singularity, by the degeneracy pressure that results from application of the quantum exclusion principle. This assumes implicitly that electrons and quarks are not fundamental particles and split into much smaller entities that generate the degeneracy pressure capable of sustaining the mass of the Universe and preventing the formation of a singularity. The model neglects the cosmological constant and yields a ground state followed by a bounce resulting in exponential inflation with no other ingredients (as dark energy) other than the fluid itself.

The Big Bang singularity represents a breakdown of current physical laws as it points to infinite density and temperature, and zero comoving volume. General relativity is one of the theories that suffer from such a breakdown: Spacetime geodesics terminate in a finite cosmological time. These and other problems (collapse of causality, loss of predictability, etc) motivate the search for non-singular cosmological models with bouncing scenarios~\cite{b1,b2,b3,b3b,b4,b5,b6,b7,b8} and the BHU is one of such scenarios.

Other efforts aiming at explaining the non-formation of the initial singularity in the Big Bang suggest that quantum mechanical effects may
temper the divergences in curvature, energy density, and temperature \cite{Ashtekar,Veneziano}.

In anticipation of a definitive and finalized theory of quantum gravity, in this work we start from the time the bounce occurred (with or without prior collapse), as in the BHU model. In Sec.~\ref{secmet} we discuss the conditions of regularity of a spherically symmetric regular spacetime (SSRS) and show that such a spacetime has no less than two horizons. In Sec.~\ref{secbhu} we discuss the BHU model and present our comments, support, and criticisms which motivates us to introduce a new FLRW patch: the Star Universe Model (SUM). In Sec.~\ref{secnov} we derive novel static spacetimes sustained by perfect fluids having constant density and pressure and determine an upper limit for the star density. In Sec.~\ref{secsum} we use the solutions derived in Sec.~\ref{secnov} to build the SUM, which is an FLRW patch with no horizons. In Sec.~\ref{secgppp} we introduce the gravitationally induced matter and entropy production formalism and discuss the matching conditions. In Sec.~\ref{secsvb} we compare BHU to SUM. In the SUM a collapse prior to a bounce is not necessary; however, we will use the term {\textgravedbl}bounce{\textacutedbl}, instead of {\textgravedbl}split{\textacutedbl}, as if a collapse occurred. We summarize in Sec.~\ref{secconc}. In this work, the metric signature is ($-,\,+,\,+,\,+$).

\section{Metric of a spherically symmetric evolving regular spacetime\label{secmet}}
Following Bondi~\cite{Bondi}, consider a spherically symmetric spacetime with metric
\begin{align}\label{m1}
&\dd s^2=-c^2 f(r,t){\rm d} t^2+\frac{\dd r^2}{g(r,t)}+r^2\dd\Omega^2,\\
&g(r,t)=1-\frac{2Gm(r,t)}{c^2r}=1-\frac{2MG\,\mathbb{D}(r,t)}{c^2r},\nonumber
\end{align} 
where $\dd\Omega^2=\dd\theta^2+\sin^2\theta \dd\varphi^2$. Here we replaced the Bondi functions ($\e ^\nu,\,\e ^\lambda$) by ($f,\,1/g$), respectively. Here $m(r,t)$ is the integral of the energy density $\rho$ within a sphere of radius $r$ with $M$ being the total ADM mass of the fluid and $\mathbb{D}(r,t)=m(r,t)/M$ is a dimensionless cumulative distribution:
\begin{align}\label{m2}
&m(r,t)=\int_0^{r}\rho(r',t)4\pi r'^2\dd r'=M\mathbb{D}(r,t),\\
&M=\int_0^{\infty}\rho(r',t)4\pi r'^2\dd r'.\nonumber
\end{align}
Note that $\rho$ is a piecewise function that vanishes beyond some expanding value $r_{\text{ex}}$ of $r$ so that the limit $\infty$ could be replaced by $r_{\text{ex}}$ (as the fluid evolves, $r=r_{\text{ex}}$ is the outer surface of the fluid). For $r\geq r_{\text{ex}}$, the spacetime~\eqref{m1} reduces to the Schwarzschild one with $f(r)=g(r)=1-\dfrac{2MG}{c^2r}$. Note that
\begin{equation}\label{rgrex}
\mathbb{D}(r)=1\quad\text{ for all }r\geq r_{\text{ex}}\,.
\end{equation}

For static or quasi-static spacetimes (the evolution time $t$ is such that the metric functions $f$ and $g$ are almost stationary), we drop the $t$ dependence from the metric functions. Now, we want that the evolving (collapsing or expanding) spherically symmetric metric~\eqref{m1} be regular, that is, be exempt from any physical singularity at the center of symmetry or elsewhere during the evolution of the mass-energy configuration. For the case of metric~\eqref{m1} to be regular, we assume that no singularity forms during evolution, that is, we assume that the energy-momentum tensor (EMT) that sources the mass configuration remains finite for all $r$ and $t$. If one of the components of the EMT diverges at some event of spacetime, the corresponding metric is non-regular.

If the static spacetime
\begin{align}\label{static}
&\dd s^2=-c^2 f(r){\rm d} t^2+\frac{\dd r^2}{g(r)}+r^2\dd\Omega^2,\\
&g(r)=1-\frac{2Gm(r)}{c^2r}=1-\frac{2MG\,\mathbb{D}(r)}{c^2r},\nonumber
\end{align}
is regular for all $r$, the density $\rho(r)$ is supposed to be finite for all $0\leq r\leq r_{\text{ex}}$ (no singular component of the EMT). For $r\ll 1$, we may replace $\rho(r')$ in~\eqref{m2} by $\rho(0)$ to obtain
\begin{equation}\label{m3}
m(r)\underset{(r\to 0)}{\simeq} \int_0^{r}\rho(0)4\pi r'^2\dd r'=
\frac{4\pi \rho(0)}{3}~r^3,
\end{equation}
yielding
\begin{equation}
g\underset{(r\to 0)}{\simeq}1-\frac{8\pi G\rho(0)}{3c^2}~r^2.
\end{equation}
Thus, any distribution with a nonvanishing value at the origin [$\rho(0)\neq 0$] yields a metric having de Sitter behavior there with an effective {\textgravedbl}cosmological constant{\textacutedbl}~\cite{smooth}
\begin{equation}\label{ic}
\Lambda=8\pi G\rho(0)/c^2.
\end{equation}

\begin{figure}[!htb]
\centering
\includegraphics[width=0.43\textwidth]{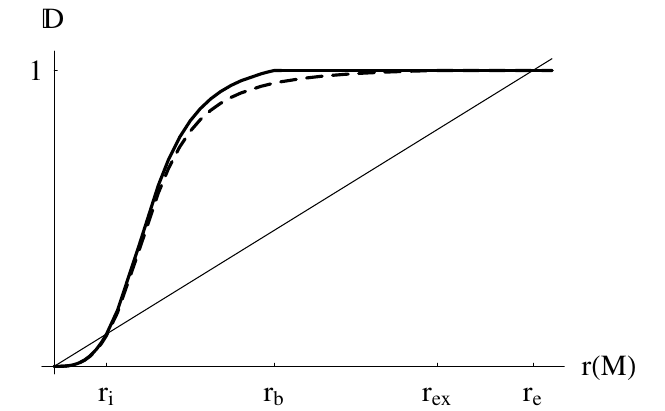} \includegraphics[width=0.43\textwidth]{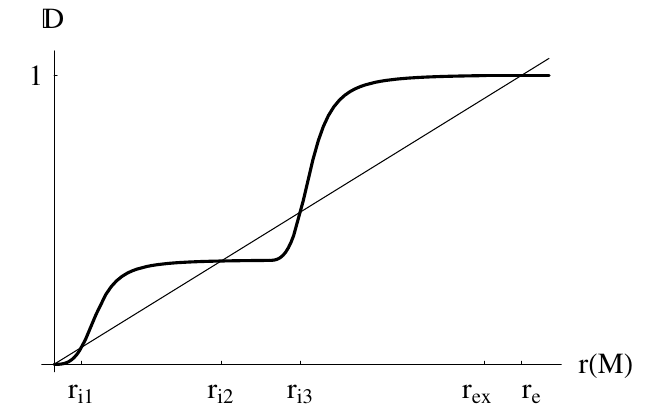} \\
\caption{\footnotesize{Plots of the line $y=c^2r/(2MG)$} and the curve $y=\mathbb{D}(r)$ (representing the cumulative distribution) for BH solution. $r_\text{e}$ event horizon, $r_\text{i1}$, $r_\text{i2}$, ..., inner horizons, $r_\text{ex}$ expanding radius (as the fluid evolves, $r=r_{\text{ex}}$ is the outer surface of the fluid), and $r_\text{b}$ bounce radius ($r_\text{b}$ is the minimum value of $r_\text{b}$). $r$ is expressed in units of $M$. Upper Panel: $\mathbb{D}(r)$ has one point of inflection (PoI) and two horizons. A bouncing evolution corresponds to a decreasing $\rho$, a state where the graph of $y=\mathbb{D}(r)$ is deformed rightward from continuous line to dashed line and so on until the graph has no intersection with the line $y=c^2r/(2MG)$; A collapsing evolution corresponds to an increasing $\rho$, a state where the graph of $y=\mathbb{D}(r)$ is deformed leftward from a situation where there is no intersection point with the line $y=c^2r/(2MG)$, to a situation where there is one intersection point, and finally to the situation shown in this panel with two intersection points. Lower Panel: $\mathbb{D}(r)$ has three PoIs and four horizons. The number of horizons is related to the number of PoIs as shown in~\cite{smooth}.}\label{Figbhu1and3}
\end{figure}

\begin{figure}[!htb]
\centering
\includegraphics[width=0.43\textwidth]{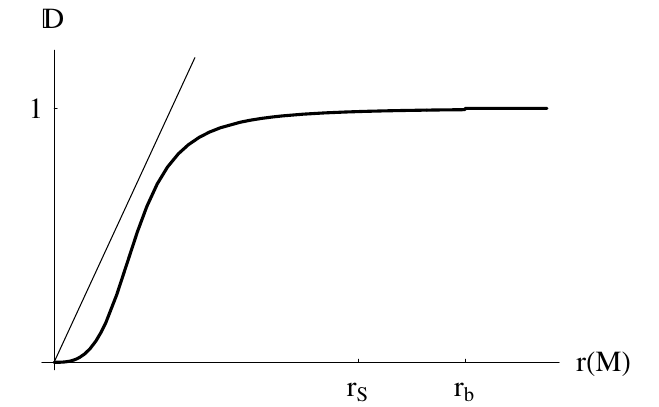}
\caption{\footnotesize{Plots of the line $y=c^2r/(2MG)$} and the curve $y=\mathbb{D}(r)$ for a star solution. In this plot $r_\text{S}=2GM/c^2$ is not a horizon.}\label{Figbhu2}
\end{figure}

Since for $r\ll 1$, $\mathbb{D}(r)=m(r)/M\to 0$ by~\eqref{m3} and for $r\geq r_{\text{ex}}$, $\mathbb{D}(r)=1$, the cumulative distribution $\mathbb{D}(r)$ has two flat sections, one for $r\ll 1$ and the other for $r\geq r_{\text{ex}}$. The shape of the graph of $\mathbb{D}(r)$ versus $r$ is the flat letter {\textgravedbl}S{\textacutedbl} if $\mathbb{D}(r)$ has one point of inflection (PoI) or two or more flat sections with many PoIs, as shown in Figs.~\ref{Figbhu1and3} and~\ref{Figbhu2}.

The horizons are solutions to equation $g(r_h)=0$, which reduces to
\begin{equation}\label{m4}
\frac{c^2}{2MG}~r_h=\mathbb{D}(r_h)\,,
\end{equation}
that is, they are the intersection of the line $y=c^2r/(2MG)$ and the curve $y=\mathbb{D}(r)$ (representing the cumulative distribution). The advantage of bringing $g(r_h)=0$ to~\eqref{m4} is that the slope of the line $y=c^2r/(2MG)$ is independent of the matter distribution $\rho$ and depends only on the total mass $M$, which is constant and so is the slope. The shape of the curve $y=\mathbb{D}(r)$ depends on the density $\rho$. As the system evolves, the shape of the line $y=c^2r/(2MG)$ remains unchanged while the graph of the curve $y=\mathbb{D}(r)$ becomes deformed leftward (if the configuration collapses) or rightward (if the configuration bounces), as shown in the upper panel of Fig.~\ref{Figbhu1and3}, 

In Fig.~\ref{Figbhu1and3} we consider the cases where $\mathbb{D}(r)$ has one PoI and two horizons (upper panel), and three PoIs and four horizons (lower panel). The number of horizons is related to the number of PoIs as shown in~\cite{smooth}. The extreme BH corresponds to the point where the line $y=c^2r/(2MG)$ is tangent to the curve $y=\mathbb{D}(r)$. Such an extreme BH is certainly unstable since, as the configuration evolves in time, the curve $y=\mathbb{D}(r)$, which depends on $\rho(r)$, changes its shape, as shown in the upper panel of Fig.~\ref{Figbhu1and3}, in such a way that it is no longer tangent to the line $y=c^2r/(2MG)$, which maintains the same shape, since it depends only on the total mass $M$.


For clarity sake, we focus on the case where $\mathbb{D}(r)$ has one PoI, as in the upper panel of Fig.~\ref{Figbhu1and3}. In this work, we assume that at $t=0$, the perfect fluid is inside the sphere $r=r_\text{b}$, with maximum density $\rho_\text{b}$, and at later times the radius of the sphere becomes $r=r_\text{ex}(t)>r_\text{b}$ (if the matter expands) with $r_\text{ex}(0)=r_\text{b}$, as shown in the upper panel of Fig.~\ref{Figbhu1and3} where $r_\text{i}$ and $r_\text{e}$ are the inner horizon and event horizon, respectively (in the lower panel,  $r_\text{i1}$, $r_\text{i2}$, ..., are inner horizons). A bouncing evolution corresponds to decreasing $\rho$ and increasing $r_\text{ex}$, a state where the graph of $y=\mathbb{D}(r)$ is deformed rightward from the continuous line to the dashed line in the upper panel of Fig.~\ref{Figbhu1and3}, and so on until the graph has no intersection with the line $y=c^2r/(2MG)$ (as in Fig.~\ref{Figbhu2}); A collapsing evolution corresponds to increasing $\rho$ and decreasing $r_\text{ex}$, a state where the graph of $y=\mathbb{D}(r)$ is deformed leftward from a situation where there is no intersection point with the line $y=c^2r/(2MG)$, to a situation where there is one intersection point (one horizon), and finally to the situation shown in this panel with two intersection points (two horizons). If c were tangent to the curve $y=\mathbb{D}(r)$ at some given time (extreme BH), this would no longer be the case shortly after a bouncing or collapsing evolution as the graph of $y=\mathbb{D}(r)$ would have been deformed. 

We have thus shown that an (evolving) \emph{regular configuration} has either no horizon, as in Fig.~\ref{Figbhu2}, or at least two horizons, as in Fig.~\ref{Figbhu1and3}. 

It is worth noticing that, in the previous arguments, $M$ represents the \emph{whole mass-energy} content in the Universe and it is constant. From the point of view of \emph{pure classical theory}, the simple graphical proof given here remains valid as far as no singularity forms and no no losses of mass-energy by radiation fluxes and similar processes occur. We do not know how quantum effects may alter the proof. However, if $M$ were the mass of a star, losses of mass to the star's environment by radiation fluxes would alter the value of $M$ by an amount $\delta M$ and the proof would hold only if $\delta M/M\ll 1$ so that the slope of the line $y=c^2r/(2MG)$ remains constant.

\section{Black Hole Universe (BHU): comments, support, and criticisms\label{secbhu}}
The BHU model applies GR to a uniform, finite FLRW fluid cloud (or patch) immersed in a Schwarzschild spacetime. Such treatments are well known in the literature and were applied in~\cite{Zeldovich,Zeldovich2,Geller,FaraoniAtieh}. 

The BHU model~\cite{GKPG} assumes a collapsing \textit{regular} configuration (an FLRW cloud), starting from some initial $r_0$ to $r=r_\text{g}$ (denoted by $R_\text{G}$ in~\cite{GKPG}), which is the radius of the collapsing cloud when it reaches the maximum density $\rho_\text{g}$ with $\rho_\text{g}'(\tau)=0$ [see right column of page 5 and Eq. (19) of~\cite{GKPG}]. In BHU, $\rho_\text{g}$ was assumed to be greater than the nuclear saturation density of the order of $\rho_{\text{nsd}}=1.4\times 10^{-13}M_\odot$ in SI units. As explained in~\cite{GKPG}: {\textgravedbl}\textit{This value of $R_\text{G}$ represents the beginning of the transition into the ground state. The model transitions from a state of constant total energy mass (with a uniform but evolving energy density) to a state of uniform and time-invariant energy density}{\textacutedbl}. The question is: Does $r_\text{g}$ remain unchanged during this transition? Based on Eq. (21) of~\cite{GKPG}, the bounce occurs at $r=r_\text{b}$ (index b is for bouncing) and this was denoted by $R_\text{B}$ in~\cite{GKPG}. It is obvious from this equation that $r_\text{b}<r_\text{g}$ because $r_\text{g}<r_\text{S}$ (according to BHU, the ground state has been formed inside the Schwarzschild radius). Contrary to the statement made in~\cite{GKPG}, we have that $\rho_\text{b}>\rho_\text{g}$, implying that $\rho_\text{b}$ is the maximum density. According to BHU, the bounce of the FLRW cloud and resulting inflation phase are driven by a quantum degenerate pressure $p=-c^2\rho$ by application of the quantum exclusion principle. As stated earlier, in this work we assume that at $t=0$, the perfect fluid is inside the sphere $r=r_\text{b}$, with maximum density $\rho_\text{b}$, and at later times the radius of the sphere becomes $r=r_\text{ex}(t)>r_\text{b}$ with $r_\text{ex}(0)=r_\text{b}$.

The whole mass of the Universe collapses to such a small radius $r_\text{b}$ is not possible within particle-physics-current knowledge, since the highest degeneracy pressure generated by application of the quantum exclusion principle can only sustain quark or other exotic stars from collapsing to form BHs. However, in the near past, nobody believed to the existence of quark stars, now we have a couple of candidates: 3C 58, XTE J1739-285, PSR B0943+10, SN 1987A, and ASASSN-15lh~\cite{quark}. From a theoretical point of view, there is the Preon Model which motivates the explanation that quarks are composed of similar preons~\cite{ELP,Hansson,HS,Horvath,Ball,Kalman1,Kalman2}. One believes that electrons are composed of preons too~\cite{Lush}. From this point of view, we support the claim made in~\cite{GKPG} that quarks and electrons are not fundamental constituents of baryonic matter, and we assume that the degeneracy pressure, generated by application of the quantum exclusion principle to these more fundamental particles, prevents the whole mass of the Universe from collapsing to form a BH.

However, the claim that the entire mass remains inside a BH with a single Schwarzschild horizon at $r_\text{S}=2GM/c^2$ is problematic~\cite{GKPG} within the classical theory as far as no singularity forms. In the graphical picture we presented in the previous section, only regular configurations, with no horizon or with at least two time-dependent horizons, form if the (bouncing/collapsing) evolution remains regular all the way. 

Consider the case of the upper panel of Fig.~\ref{Figbhu1and3}, where there are exactly two horizons. Now, let us assume, as in~\cite{GKPG}, that the whole mass of the ground state is inside a Schwarzschild radius $r_\text{S}$, that is, $r_\text{b}<r_\text{S}=2MG/c^2$, where $r_\text{b}$ is the radius of the ground state (before the bounce took place). There are three cases. 
\begin{enumerate}
\item The case $r_\text{b}<r_\text{i}$ is not possible because $\mathbb{D}(r_\text{b})=1$ (at $t=0$ the whole mass $M$ is inside the sphere $r=r_\text{b}$). If this case were possible, the graph of $y=\mathbb{D}(r)$ would reach the value 1 before crossing the line $y=c^2r/(2MG)$.
\item If $r_\text{i}<r_\text{b}<r_\text{e}=r_\text{S}$, how did the formation of a ground state proceed during collapse if $r$ is timelike for $r_\text{i}<r<r_\text{e}$? During collapse, after the particles cross $r_\text{e}$, they will continue to move to decreasing $r$ until, normally, they cross $r_\text{i}$~\cite{HK}. The collapse proceeds as follows (see the upper panel of Fig.~\ref{Figbhu1and3}). At the beginning of collapse, the Universe behaves as a star and the graph of $y=\mathbb{D}(r)$ is entirely on the right of the line $y=c^2r/(2MG)$ with no intersection point (no horizon as in Fig.~\ref{Figbhu2}). As the collapse progresses, the density increases and the graph of $y=\mathbb{D}(r)$ deforms until a tangent point forms with the line $y=c^2r/(2MG)$ (a horizon is formed). Then, this horizon splits into two as the graph of $y=\mathbb{D}(r)$ continues to deform leftward (formation of two horizons).

This case is problematic from another point of view. If the bounce occurs before a total collapse ($r_\text{i}<r_\text{b}<r_\text{e}$), knowing that the coordinate $r$ is spacelike for $0<r<r_\text{i}$, the bouncing particles of the perfect fluid from this region ($0<r<r_\text{i}$) will cross the surface at $r=r_\text{i}$ in the sense of increasing $r$ and, since in the region $r_\text{i}<r<r_\text{e}$ the coordinate $r$ is timelike, these particles will continue to move to increasing $r$ until they cross the event horizon $r_\text{e}$. Thus, the surface at $r=r_\text{e} (=r_\text{S})$ is not a cosmological horizon, as claimed in~\cite{GKPG}. There is a similar criticism in~\cite{HK} and even if we assume that $r_\text{i}\equiv 0$, which is the case where one horizon only would have formed (by some quantum process), the criticism persists as $r$ would have become timelike for all $r\in [0,\,r_\text{e}]$. This criticism is purely from a classical point of view and there is no formalism in quantum theory to describe regions with horizons.
\item The BHU model assumes that the fluid first collapsed from some initial state of mater distribution (with some energy budget) in such a way that some amount of the fluid was outside the Schwarzschild radius $r_\text{S}$, which is constant and depends only on the total mass, and another amount was inside of it. However, after the bounce occurred, the total mass remained within the radius $r_\text{S}$, as claimed in~\cite{GKPG}. Drawback: the fluid does not find its initial state of matter distribution, that is, the region outside the sphere $r=r_\text{S}$ is no longer accessible to the fluid.
\end{enumerate}

In the subsequent sections, we will build a novel Universe model assuming that the collapse ended while the graph of $y=\mathbb{D}(r)$ is still entirely on the right of the graph of the line $y=c^2r/(2MG)$ (no intersection point -- no horizon), that is, before the formation of any horizon, so that the ground state is a star with $r_\text{b}>r_\text{S}$. This is the Star Universe Model (SUM), which is not a substitute to BHU; rather, the model departs from the idea of collapse with horizons, assumes that the degeneracy pressure, generated by application of the quantum exclusion principle, halts the collapse prior to the formation of horizons. It leads to new conclusions and results.

\section{Static solutions with constant density and pressure\label{secnov}}
In this section, we will present two SSRSs sourced by perfect fluids that have constant density and pressure (with no cosmological constant). The solutions will be different from the interior Schwarzschild metric (Schwarzschild fluid solution)~\cite{SF,Tolman}, which is sourced by perfect fluid of constant density and decreasing pressure away from the center of the star.

The metric is of the form~\eqref{m1} with no dependence on $t$. The field equations $\mathcal{F}^{\mu\nu}:=G^{\mu\nu}-\kappa T^{\mu\nu}=0$ with $\kappa =8\pi G/c^4$, $T^{\mu\nu}=(\rho +p/c^2)u^\mu u^\nu +pg^{\mu\nu}$, $u^\mu=\big(1/\sqrt{f},\,0,\,0,\,0\big)$, $R^{\alpha}{}_{\beta\mu\nu}=\Gamma^{\alpha}{}_{\beta\nu,\mu}-\cdots$ and $R_{\beta\nu}=R^{\alpha}{}_{\beta\alpha\nu}$ take the form assuming that $\rho$ and $p$ are constants (uniform)
\begin{align}
\label{n1}&\mathcal{F}^{tt}=0:\; m'(r)=4\pi\rho r^2 \Rightarrow m(r)=\frac{4\pi\rho }{3}~r^3\,,\\
\label{n2}&\mathcal{F}^{rr}=0:\; -\frac{2 G m(r)}{c^2 r^3}+\frac{[c^2 r-2 G m(r)] f'(r)}{c^2 r^2 f(r)}=\kappa  p\,.
\end{align}
In this paper, the prime notation means derivative with respect to the argument shown between parentheses. For instance, $m'(r)$ is $\dd m/\dd r$, $a'(\tau)$ is $\dd a/\dd \tau$, and $a''(\tau)$ is $\dd^2 a/\dd \tau^2$. Knowing $m(r)$ from~\eqref{n1} we can obtain the general expression of $f(r)$. The first solution is eminent and it corresponds to $f=\text{const}$, $p=-c^2\rho /3$ and $m(r)=4\pi\rho r^3/3$. Curiously enough, this has never ever been discussed in the literature as it is different from the de Sitter solution. Analytically speaking, this was derived in the context of Einstein Static Universe\footnote{Rigorously speaking, Einstein solution corresponds to a dust $\rho$ (no pressure) with a cosmological constant $\Lambda=\kappa\rho/2$. This also is obtained from  Eq.~(4.1) of~\cite{Tolman} upon setting $\Lambda=1/R_E^2$ with $R_E$ being the radius of curvature~\cite{1917}.} in Eq.~(4.1) of~\cite{Tolman}, which when setting $\Lambda =0$ reduces to $p=-c^2\rho /3$. However, in~\cite{Tolman} the cosmological constant $\Lambda$ was supposed to be positive and bounded below and above by some positive constant and its third multiple to ensure the positiveness of $\rho$ and $p$. The second solution is, of course, the de Sitter solution. Here we are solving the field equation assuming no cosmological term but just a matter term in the field equation. In this work there will be no notion of cosmological constant and the one given in~\eqref{ic} is for purposes of comparison.

The two static solutions corresponding to $p=-c^2\rho$ and $p=-c^2\rho /3$ take the form, respectively,
\begin{multline}\label{s1dS}
\dd s^2=-c^2 \Big(1-\frac{8\pi G\rho r^2}{3c^2}\Big){\rm d} t^2+\frac{\dd r^2}{\Big(1-\frac{8\pi G\rho r^2}{3c^2}\Big)}+r^2\dd\Omega^2,
\end{multline}
[which yields the same constant as in~\eqref{ic}] and
\begin{multline}\label{s2dS}
\dd s^2=-c^2 \Big(1-\frac{2 GM}{c^2r_\text{ex}}\Big){\rm d} t^2+\frac{\dd r^2}{\Big(1-\frac{8\pi G\rho r^2}{3c^2}\Big)}+r^2\dd\Omega^2\,.
\end{multline}

\section{Star Universe Model (SUM)\label{secsum}}
In the scientific literature there is a long and apparently an {\textgravedbl}open{\textacutedbl} list of bouncing scenarios~\cite{b1,b2,b3,b3b,b4,b5,b6,b7,b8}, the recent of which is the BHU. In these models the bounce is assumed to occur due to violation of the strong energy condition, to a repulsive force introduced by quantum corrections to the field equations, or to exotic scalar fields with negative kinetic energy, depending on the model. The BHU applied the FLRW patch model~\cite{Zeldovich,Zeldovich2,Geller} and assumed that the bounce occurred inside a BH due to a degenerate negative pressure. In the SUM we will also apply the FLRW patch model to a matter-energy configuration (with no horizon) which initially (just prior to the bounce) has already a positive spatial curvature, as described in the following paragraphs. The evolving spherical symmetric metric~\eqref{m1} is described as an FLRW cloud or patch~\cite{Zeldovich,Zeldovich2,Geller}, the so-called FLRW analogy~\cite{FaraoniAtieh}. The \emph{finite} FLRW cloud is assumed to have the same symmetrical properties of the FLRW spacetime itself.

As we wrote in the Introduction, awaiting a definitive and finalized theory of quantum gravity, the \emph{determination of the quantum state} of the fluid confined into a sphere of radius $r_\text{b}$, with total mass $M$, remains beyond current physics knowledge. All cosmological models, including the Big Bang model itself, assume that the matter and energy content of the Universe were torn apart at some initial time $\tau =0$. The Big Bang model relies on Hawking-Penrose singularity theorems, it assumes that the whole mass of the Universe was confined into a point (with no dimensions), and it does not present any statistical model by which the partition function along with $p=-c^2\rho$, of the initial quantum state, are determined. The SUM relies on~\eqref{s1dS}, which is a solution to the field equations, it assumes that the whole mass of the Universe was confined into a sphere of radius $r_{\text{b}}>0$, and it does not present any statistical model by which the EoS $p=-c^2\rho$, of the initial quantum state, is determined. The initial quantum state of the Universe is ignored in all bouncing models. In inflationary models, to by-pass the difficulty, a homogeneous scalar (classical) field with energy density $[\phi'(\tau)+2V(\phi)]/2$ and pressure $[\phi'(\tau)-2V(\phi)]/2$~\cite{Mukhanov} was introduced\footnote{There are other alternatives as the two measures field theory, which introduces two scalar fields and allows for inflation~\cite{Campo}, and the power-law plateau inflationary model~\cite{Jawad2}.} to justify the relation $p=-c^2\rho$, however, this is nothing but a way to express our lack of knowledge of the initial quantum state of the fluid confined into a sphere of radius $r_\text{b}$. It is clear that such an approach is not sufficient to explore all properties of the quantum state, however, it has removed the speculative character of $p=-c^2\rho$ in the Big Bang theory since scalar fields are no longer hypothetical and their existence has been experimentally proven in the Higgs field model.

It is well known that the spacetime that describes the de Sitter Universe has three representative metrics: closed, open, and flat~\cite{Mukhanov}. We are interested in the closed form
\begin{equation}\label{sum1}
\dd s^2=-c^2\dd\tau^2 + \ell^2\cosh^2\Big(\frac{c\tau}{\ell}\Big)(\dd \xi^2 +\sin^2\xi\dd\Omega^2)\,,
\end{equation}
where $\ell$ is called the curvature scale or the radius of the hyperboloid of one sheet embedded in 4-dimensional Minkowski space and, in cosmology, is usually introduced by $H_\Lambda=c/\ell$ and $\Lambda=3/\ell^2$. The static coordinates ($t,\,r$) are related to FLRW coordinates by
\begin{align}\label{sum2}
&\sqrt{f(r)}\sinh\Big(\frac{ct}{\ell}\Big)=\sinh\Big(\frac{c\tau}{\ell}\Big)\,,\nonumber\\
&\sqrt{f(r)}\cosh\Big(\frac{ct}{\ell}\Big)=\cosh\Big(\frac{c\tau}{\ell}\Big)\cos\xi\,,\\
&r=\ell\sin\xi\cosh\Big(\frac{c\tau}{\ell}\Big)\,,\qquad f(r)=1-\frac{r^2}{\ell^2}\,.\nonumber
\end{align}
In static coordinates, the metric, $\dd s^2=-c^2f(r)\dd t^2+\dd r^2/f(r)+r^2\dd\Omega^2$, is of the form~\eqref{s1dS} with
\begin{equation}\label{sum3}
\ell^2 =\frac{3c^2}{8\pi G\rho}\,,\qquad H_\Lambda =\sqrt{\frac{8\pi G\rho}{3}}\,.
\end{equation}

In our Star Universe Model (SUM), the dynamics of the de Sitter Universe is not dominated by a cosmological constant or inflaton; rather, by an isotropic perfect fluid endowed with the highest negative degeneracy pressure $p=-c^2\rho$ capable of halting the collapse of the fluid to a singularity. There is no statistical model that leads to $p=-c^2\rho$. As we shall see in Sec.~\ref{secgppp}, $p$ is not the thermodynamic pressure of the fluid; rather, it is a negative pressure resulting from a gravitationally induced matter and entropy production process. We do not assume that the quantum state of the fluid confined into a sphere of radius $r_\text{b}$ is necessarily the result of a prior collapse of its matter content, but this is not excluded, however, we will continue to use the term {\textgravedbl}bounce{\textacutedbl}, instead of {\textgravedbl}splitting{\textacutedbl}, as if it were a collapse prior to the formation of a ground state. 

As we explained earlier, there is in the literature an explanation to the state $p=-c^2\rho$ via the introduction a homogeneous scalar (classical) field with energy density $[\phi'(\tau)+2V(\phi)]/2$ and pressure $[\phi'(\tau)-2V(\phi)]/2$~\cite{Mukhanov}, however, this is nothing but a way to express our lack of knowledge of the quantum state of the fluid confined into a sphere of radius $r_\text{b}$. It is clear that such an approach is not sufficient to explore all properties of the quantum state. This practice is well known in physics: a scalar function $\phi(x)$ was introduced to explain the propagation of light along the $x$-axis, and later it was discovered that $\phi(x)$ is just one of the components of the electric field ($E_y$ or $E_z$)~\cite{Buchwald}. In our SUM, we do not introduce any scalar field or cosmological constant; all we assume is that the fluid particles were tightly sewn together in a ground state sourcing an SSRS defined by~\eqref{s1dS}, \eqref{s3}, \eqref{s4} and $p=-c^2\rho$, then the bounce (splitting), caused by the negative pressure (anti-gravity effect), tore them apart at some initial time that we take $\tau =0$. Since the spacetime describing the fluid has no horizon (Fig.~\ref{Figbhu2}), the fluid particles make up a star rather than a BH.

Notice that the drawback discussed in the BHU model is not encountered in the SUM: if a collapse occurred, then after the bounce the fluid would find its initial state just before collapse with the same energy budget.

To move further, we assume that GR remains valid at such small radii (application of its extension $f(R)$ gravity, with $R$ being the Ricci scalar, does not change qualitatively the subsequent results). We also assume that the hypotheses of homogeneity and isotropy remain valid during the bounce, but the fluid-particles range remains finite; that is , we assume that the description of spacetime events within the cloud follows that of closed FLRW as in the FLRW patch model~\cite{Zeldovich,Zeldovich2,Geller}. Since our initial metric at $\tau =0$ (defined by~\eqref{s1dS}, \eqref{s3}, \eqref{s4} and $p=-c^2\rho$) \emph{has already a constant positive spatial curvature $k$, we assume that the bounce did not affect the value of $k$.} To this end, we transform~\eqref{sum1} by introducing a new co-moving coordinate $\chi$ defined by
\begin{align}\label{sum4}
&\sin\xi =\sqrt{k}\,\chi \,,\qquad \ell_\text{n}=\ell\sqrt{k}\,,\nonumber\\
&a(\tau)=\ell_\text{n}\cosh\Big(\frac{c\tau}{\ell}\Big)\,,\qquad r=a\chi\,,
\end{align}
to obtain
\begin{equation}\label{sum5}
\dd s^2=-c^2\dd\tau^2 + a(\tau)^2\,\frac{\dd \chi^2}{1-k\chi^2} +a(\tau)^2\chi^2\dd\Omega^2\,,
\end{equation}
and this is the metric we will use for the subsequent investigation. The FLRW field equations take the form
\begin{align}
\label{eq1}&\frac{a'(\tau )^2}{a(\tau )^2}=\frac{8 \pi  G \rho (\tau )}{3}-\frac{c^2 k}{a(\tau )^2}\,,\\
\label{eq2}&\frac{a''(\tau )}{a(\tau )}=-\frac{4 \pi G}{3c^2}~[c^2 \rho (\tau )+3p(\tau )]\,,\\
\label{eq3}&c^2 \rho '(\tau )+3 \frac{a'(\tau )}{a(\tau )} [c^2 \rho (\tau )+p(\tau )]=0\,,	
\end{align}
where the prime denotes the derivative with respect to the coordinate shown in parentheses. Since we assumed to be valid the homogeneity hypothesis, the density and pressure are uniform. 

\section{Gravitationally induced matter and entropy production -- matching conditions\label{secgppp}}
Starting from now on we will distinguishing between thermodynamic pressure $p$ and total pressure $p_{\text{tot}}$. In a non-evolving scenario, both pressures are equal, as in Sec.~\ref{secnov}. In the case of particle production the two pressure are not equal.

In the gravitationally induced matter and entropy production formalism~\cite{GPPP1,GPPP2}, matter is usually described by a perfect fluid with an energy-momentum tensor (EMT)  given by
\begin{equation}\label{emt}
T^{\mu\nu}=\Big(\rho+\frac{p}{c^2}\Big)u^\mu u^\nu + pg^{\mu\nu}\,,
\end{equation}
where $u^\mu$ is the normalized four-velocity of the comoving fluid element ($u^\mu u_\mu =-c^2$), $p$ is the \emph{thermodynamic} pressure and $\rho$ is the energy density both measured in a localized inertial frame comoving with the fluid.

Matter and entropy are produced during the first evolution stage of the universe~\cite{GPPP1}. This process results in a negative pressure $p_c$, other than the thermodynamic pressure $p$, the value of which is determined by applying the laws of thermodynamics~\cite{GPPP1} applied to nonequilibrium states~\cite{GPPP2} (and references therein)
\begin{equation}\label{p4}
p_c(\tau)=-\frac{\Gamma(\tau)[c^2\rho(\tau)+p(\tau)]}{3}~\frac{a(\tau)}{\dot{a}(\tau)}\,,
\end{equation}
where $\Gamma(\tau)>0$ is the homogeneous particle production rate\footnote{Very similar expression for the production rate was derived in~\cite{dark} when dark energy was considered as an effective manifestation of nonequilibrium thermodynamics [see Eqs.~(17)-(18) of~\cite{dark}].}. In this matter and entropy production formalism, the entropy per particle $\sigma$ is assumed constant during evolution. If $n(\tau)$ denotes the homogeneous particle number density, $s(\tau)$ the homogeneous entropy density, and $\rho(\tau)$ the homogeneous mass-energy density, then $\sigma=s/n$ and the balance equations for nonequilibrium states take the form~\cite{GPPP1,GPPP2}
\begin{align}
&n'(\tau)+3~\frac{a'(\tau )}{a(\tau )}~n=\Gamma n\,,\nonumber\\
\label{p1}&s'(\tau)+3~\frac{a'(\tau )}{a(\tau )}~s=\Gamma s \,,\\
&\rho'(\tau)+3~\frac{a'(\tau )}{a(\tau )}~\Big(\rho+\frac{p}{c^2}\Big)=\Gamma \Big(\rho+\frac{p}{c^2}\Big)\,,\nonumber
\end{align}
where $a(\tau)$ is the scale factor~\eqref{sum5}.

In terms of $\Gamma$, the field equations take the form~\cite{GPPP2,GPPP3}
\begin{align}
&\frac{a'(\tau )^2}{a(\tau )^2} = \frac{8\pi G}{3}~\rho-\frac{c^2k}{a^2}\,,\nonumber\\
&\frac{a''(\tau )}{a(\tau )} = -\frac{4\pi G}{3} \Big(\rho+\frac{3p}{c^2}\Big) + \frac{4\pi G}{3}\Gamma \Big(\rho+\frac{p}{c^2}\Big)~\frac{a(\tau )}{a'(\tau )}\,,\nonumber\\
\label{p5}&\rho'(\tau)+3~\frac{a'(\tau )}{a(\tau )}~\Big(\rho+\frac{p}{c^2}\Big)=\Gamma \Big(\rho+\frac{p}{c^2}\Big)\,,
\end{align}
which reduce to FLRW equations~\eqref{eq1}-\eqref{eq3} when the gravitationally induced matter and entropy production ceases.

Note that Eqs.~\eqref{p5} have equivalent form in terms of the total pressure $p_{\text{tot}}\equiv p+p_c$:
\begin{align}
\label{p6m1}&\frac{a'(\tau )^2}{a(\tau )^2} = \frac{8\pi G}{3}~\rho-\frac{c^2k}{a^2}\,,\\
\label{p6m2}&\frac{a''(\tau )}{a(\tau )} = -\frac{4\pi G}{3} \Big(\rho+\frac{3p_{\text{tot}}}{c^2}\Big)\,,\\
\label{p6}&\rho'(\tau)+3~\frac{a'(\tau )}{a(\tau )}~\Big(\rho+\frac{p_{\text{tot}}}{c^2}\Big)=0\,,
\end{align}
which means that the effective EMT is that of a perfect fluid with pressure $p_{\text{tot}}$ and energy density $\rho$~\cite{GPPP2}:
\begin{equation}\label{emteff}
T^{\mu\nu}=\Big(\rho+\frac{p_{\text{tot}}}{c^2}\Big)u^\mu u^\nu + p_{\text{tot}}g^{\mu\nu}\,.
\end{equation}

\subsection{Matching conditions}
We assume that the Universe exists initially at $\tau=0$ in a de Sitter state defined by~\eqref{s1dS} with a total pressure satisfying $p_{\text{tot}}=-c^2\rho$. This is the state with the highest negative pressure (capable of preventing collapse). The corresponding external metric is that of Schwarzschild and the matching takes place at the radius of the ground state $r=r_\text{b}$, which marks the surface of the star at $\tau=0$. As we shall see later, the bounce will cause the scaling factor to inflate exponentially. Matching the de Sitter solution~\eqref{s1dS} with Schwarzschild line element, yields the matching constraint:
\begin{equation}\label{s2}
1-\frac{8\pi G\rho r_\text{b}^2}{3c^2}=1-\frac{2GM_{\text{g}}}{c^2r_\text{b}}\,,
\end{equation}
where $M_{\text{g}}$ is the gravitational mass and $\rho_{\text{b}}=\rho$. This leads to
\begin{equation}\label{s3}
r_\text{b}=\Big(\frac{3M_{\text{g}}}{4\pi\rho}\Big)^{1/3}\,,
\end{equation}
and we assume $r_\text{b}>r_\text{S}=2GM_{\text{g}}/c^2$, that is,
\begin{equation}\label{s4}
\rho <\frac{3 c^6}{32 \pi  G^3 M_{\text{g}}^2}\Leftrightarrow \rho M_{\text{g}}^2<\frac{3 c^6}{32 \pi  G^3}\Leftrightarrow \frac{(MM_{\text{g}}^2)^{1/3}}{r_\text{b}}<\frac{c^2}{2G}\,,
\end{equation}
where $M=4\pi\rho r_\text{b}^3/3$ is the cumulative mass-energy parameter of the Universe\footnote{The inertial mass $M_{\text{inertial}}$ is defined by \[M_{\text{inertial}}=4\pi\rho\int_0^{r_{\text{b}}} \frac{x^2\dd x}{\sqrt{1-\frac{x^2}{\ell^2}}}\,,\qquad \ell^2 =\frac{3c^2}{8\pi G\rho}\,. \] This yields \[M_{\text{inertial}}=4\pi\rho \Big[\frac{\ell^3}{2}\arcsin\Big(\frac{r_{\text{b}}}{\ell}\Big)-\ell r\sqrt{\ell^2-r_{\text{b}}^2}\,\Big]\,, \] which reduces to $M=4\pi\rho r_\text{b}^3/3$ for small $r_{\text{b}}$. Using $\xi =\arcsin(r_{\text{b}}/\ell)$ and $a_0\equiv a(\tau=0)$ as defined in~\eqref{sum4} and~\eqref{sum5}, we obtain \[M_{\text{inertial}}=4\pi\rho \frac{a_0^3}{k^{3/2}}\Big(\frac{\xi}{2}-\frac{\sin 2\xi}{4}\Big)\,,\] which is the same as that given in~\cite{Zeldovich2} where $k$ was set equal to $1$. The gravitational mass $M_\text{g}$ is the same mass $M$ introduced in Sec.~\ref{secmet}, which is conserved (as it is the total energy content in the Universe) since the production process only converts gravitational energy to mass.} Knowing its gravitational mass, the first expression in~\eqref{s4} determines the largest density a Universe may have. In SUM we also assume that $\rho_{\text{b}}>\rho_{\text{nsd}}$ (the nuclear saturation density).

As is well known the gravitational mass of closed or semi-closed Universe is different from its inertial mass due to gravitational defect~\cite{Zeldovich2}; for a pressure-less FLRW cloud, the ratio of the two masses depends on the comoving coordinate $\chi$. In our model the Universe is semi-closed because it does not emerge from a singularity where at the initial time the scaling factor $a=0$ and $r=0$; rather, at the initial time we have $r=r_\text{b}\neq 0$.

Now, using the constraint~\eqref{s4}, the graph of $y=\mathbb{D}(r)$ will have the same shape as that of Fig.~\ref{Figbhu2} with an infinite plateau for $r\geq r_\text{b}$ and will have no intersection with the line $y=c^2r/(2MG)$.

We assume that \emph{initially} during the production process the thermodynamic pressure $p\equiv 0$ and, since in the external Schwarzschild region the pressure is 0, there is no further constraint to impose to ensure full matching of the internal and external solutions. 

As the expansion proceeds, the thermodynamic pressure increases ($p\neq 0$ and $p>0$) due to production of mass, entropy , and particles, it is no longer possible to match the evolving metric~\eqref{sum5} [which at $\tau=0$ was given by~\eqref{s1dS}] with an external Schwarzschild metric. The mass and particle production modifies the surrounding of the star, in this case the evolving internal metric~\eqref{sum5} is matched to the external generalized Vaidya metric,
\begin{equation}\label{Vaidya}
\dd s_+^2=-c^2\Big(1-\frac{2GM(r_v,v)}{c^2r_v}\Big)\dd v^2 -2c\dd v\dd r_v  + r_v^2\dd\Omega^2\,.
\end{equation}
Details of the matching conditions do not depend on whether the evolving internal metric~\eqref{sum5} is collapsing or expanding and are given in chapter 3 of~\cite{junction} and in~\cite{junction2,junction3} and references therein. The matching procedure is outlined in the Appendix~\ref{secaa}.

\subsection{Bounce\label{secbounce}}

To determine $a(\tau)$ during the first phase of mass production ($\tau\gtrsim 0$), we solve~\eqref{p4} with $p=0$ and $p_c=p_{\text{tot}}=-c^2\rho$, that is, we assume that initially the Universe is in the de Sitter state~\eqref{s1dS} and that all its content is in the form of pure energy, which is getting converted to mass (particles) at the rate $\Gamma(\tau)$. We find
\begin{equation}\label{f1}
a(\tau)=a_0\exp\Big(\frac{1}{3}\int_{}^{\tau}\Gamma(u)\dd u\Big)\,.	
\end{equation}
Equations~\eqref{p1} lead to $\dot{n}=\dot{s}=\dot{\rho}=0$, and the second line in~\eqref{p5} yields $\ddot{a}/a=8\pi G\rho/3$, and since $\rho$ is constant we obtain
\begin{equation}\label{f2}
a(\tau)=a_0\exp(H_\Lambda\tau)\,,
\end{equation}
where $H_\Lambda^2\equiv 8\pi G\rho/3$~\eqref{sum3}. On comparing~\eqref{f1} and~\eqref{f2}, we obtain $\Gamma =3H_\Lambda (=c^2\Lambda)$, and this constant value of $\Gamma$ solves the first line in~\eqref{p5} with $k=0$. Thus, the smooth expansion or big bang starts with constant rate of production, constant density and entropy ($=0$), and exponential expansion. All that remains true at the very initial time $\tau=0$.

Thus, the \textit{birth of the Universe was inflationary by the laws of GR alone and the assumption that initially $p_c=-c^2\rho$ along with the homogeneity and isotropy hypotheses}. The assumption that initially $p_c=-c^2\rho$ is not rejected by the laws of classical GR and its extensions, as it yields an SSRS~\eqref{s1dS} that is a solution to the field equations.

A smooth graceful exit from inflationary state might be phenomenologically described as follows. We may assume that a description of our Universe by a de Sitter one lasted $10^{-32}$ s (as in $\Lambda$CDM model), a period during which the density reduces by some factor and the absolute value of the negative pressure has decreased by a similar factor but remains negative. As the Universe expands, both entities, $\rho$ and $|p_{\text{tot}}|$, continue to decrease until they meet the constraint $p_{\text{tot}}=-c^2\rho /3$ of the other static solution (generalized static Einstein Universe)~\eqref{s2dS} at some time $\tau_\text{E}$ when the scale factor, density and pressure were $a_\text{E}$, $\rho_\text{E}$ and $p_\text{E}$, respectively. The details of the evolution of the Universe from $\tau=10^{-32}$ s to $\tau_\text{E}$ could be handled with if a model for an equation of sate (EoS) were available. For $\tau=\tau_\text{E} \pm \Delta \tau$, since~\eqref{s2dS} is a static solution, one may assume that the constraint $p_{\text{tot}}=-c^2\rho /3$ remains quasi satisfied for a short period of time. In this case the second line in~\eqref{p6} implies a linear expansion:
\begin{equation}\label{eq7}
a(\tau)=c_\text{E}(\tau-\tau_\text{E})+a_\text{E}\,, \quad (c_\text{E}>0)\,.
\end{equation}
Now, the third line in~\eqref{p6} reduces to $\rho'(\tau)/\rho=-2a'(\tau)/a$ and implies
\begin{equation}\label{eq8}
\rho =\frac{\rho_\text{E}a_\text{E}^2}{a^2}\,, 
\end{equation}
and similar expression for pressure: $p_{\text{tot}}=p_\text{E}a_\text{E}^2/a^2$ with $p_\text{E}=-c^2\rho_\text{E} /3$. To our knowledge, these expressions for $\rho$ and $p$ in term of $a$ or $\tau$ are not available in the literature. Why does $\rho$ not follow the law $a^{-3}$? In this linear regime~\eqref{eq7}, both the thermodynamic pressure $p\neq 0$ (since $n$ is no longer 0 because of mass and particle production) and $p_c\neq 0$ with $p_{\text{tot}}<0$, the increasing negative pressure $p_{\text{tot}}$ (approaching 0) is converted to energy density due to mass production: as the Universe expands, repulsive anti-gravity force is turning off and attractive gravity force is turning on, thus generating more potential energy.

As we have seen, the total pressure in absolute value decreases faster than the density. This regime is likely to continue until both entities are positive. Since there is no static solution to the field equations sourced by dust (with constant or variable density), $p_{\text{tot}}=0$, there is no special behavior for $a(\tau)$ when $p_{\text{tot}}$ vanishes (and $\rho >0$). The phases light-dominant, matter-dominant, etc, occur in subsequent periods of evolution qualitatively as in the $\Lambda$CDM model. Roughly speaking, $a(\tau)$ evolves as $\propto\exp(H_\Lambda\tau)$, $\propto c_\text{E}(\tau-\tau_\text{E})+a_\text{E}$, $\propto\tau^{1/2}$ (light-dominant), $\propto\tau^{2/3}$ (matter-dominant). The lacking behavior is between inflation and linear evolution and between linear and light-dominant evolution.

\subsection{Transformation from co-moving frame to rest frame}
Both BHU and SUM introduce a static observer sitting in the rest frame with origin coinciding with the center of the spherical symmetry. We want to relate the observations made by the static observer~\eqref{m1} to those made by the co-moving frame with the fluid~\eqref{sum5}, with the aim to apply them to the SUM. This work was done in~\cite{GKPG} and corrections are needed before we can move forward: we will introduce corrections to Eq.~(50) of~\cite{GKPG} (this equation was also given in~\cite{EKI}), as well as other equations in~\cite{GKPG}, then apply the corrected formulas to SUM. In the following, the matrix $\Lambda$
\begin{equation}\label{mat}
\Lambda=\begin{pmatrix}
	\partial_\tau t & \partial_\chi t  \\
	\partial_\tau r & \partial_\chi r
\end{pmatrix}\,,
\end{equation}
transforms coordinates $(\tau,\chi)$ in the co-moving frame to coordinates $(t,r)$ in the rest frame. Here, the rest frame is attached to the center of the configuration. We first determine the transformation matrix $\Lambda$ for the de Sitter Universe, where $f=g=1+2\Psi=1+2\Phi$ (in the notation of~\cite{GKPG}). Here, $f=g$ is given in~\eqref{s1dS}. We obtain $\partial_\tau t=\sqrt{1-k\chi^2}/f$,  $\partial_\chi t=arH /(c^2f\sqrt{1-k\chi^2})$, $\partial_\tau r=r H$ and $\partial_\chi r=a$, with $r=a\chi$, $H=a'/a$ and $a'a=Ha^2$. These derivatives are directly obtained from~\eqref{sum2}, \eqref{sum4} and~\eqref{sum5} and correspond to $\tau =0$. During expansion ($\tau >0$), and again using the notation in Eq.~(46) of~\cite{GKPG}, we have $f(r,t)=1+2\Psi$ and $g(r,t)=1+2\Phi$ [and $\Phi<0$~\eqref{m1}], we obtain the elements of the matrix $\Lambda$ as: 
\begin{align}
&\partial_\tau t=\frac{\sqrt{1-k\chi^2}}{\sqrt{fg}}\,,\quad \partial_\chi t =\frac{a\sqrt{-2\Phi-k\chi^2}}{c\sqrt{fg}\sqrt{1-k\chi^2}}\,,\nonumber\\
\label{trans}&\partial_\tau r=c\sqrt{-2\Phi-k\chi^2}\,,\quad \partial_\chi r=a\,,.
\end{align}
Notice the presence of the factor $\sqrt{1-k\chi^2}$ in $\partial_\tau t$ and its absence in $\partial_\chi r$. These expressions introduce corrections to those given in~\cite{GKPG}.

For $\tau >0$, inspired by the de Sitter case, we can take\footnote{Since $r=a\chi$, we have $\partial_\tau r=a'(\tau)\chi =Ha\chi =Hr$. Note that if~\eqref{trans} are satisfied, then~\eqref{m1} implies~\eqref{sum5}. With $c\sqrt{-2\Phi-k\chi^2}=rH$, the integrability condition $\partial_{\chi\tau}r=\partial_{\tau\chi}r$ is satisfied. The other integrability condition $\partial_{\chi\tau}t=\partial_{\tau\chi}t$ is also satisfied provided $\{a(\tau),\,f(a(\tau)\chi),\,g(a(\tau)\chi)\}$ are subject to some differential equation to set of solutions of which is non-empty.} $\partial_\tau r =rH=u$ ($u$: velocity of an element of the fluid with respect to the rest frame), that is, $c\sqrt{-2\Phi-k\chi^2}=rH$, and this yields $2\Phi=-(r^2H^2/c^2)-k\chi^2$, which is different from the expression given in~\cite{GKPG}. This affects the expressions in Eq.~(53) of~\cite{GKPG} if $k\neq 0$:
\begin{equation}\label{mat2}
\bar{\rho }=\frac{\beta ^2 p_{\text{tot}}+c^2 (1-k \chi ^2) \rho }{c^2 (1-\beta ^2-k \chi ^2)},\quad \bar{p}_{\text{tot}}=\frac{c^2 \beta ^2 \rho +(1-k\chi ^2) p_{\text{tot}}}{1-\beta ^2-k \chi ^2}\,,
\end{equation}
where $\beta =u/c=rH/c$. $(\rho,p_{\text{tot}})$ are measured in the co-moving frame and $(\bar{\rho },\bar{p}_{\text{tot}})$ are the corresponding entities in the rest frame.

\subsection{Possible observable consequences}

At $\tau=0$ the two frames have relative velocity $u =0$, it is trivial to obtain $\bar{p}_{\text{tot}}=-c^2\bar{\rho}=p_{\text{tot}}=-c^2\rho$. It may be possible during expansion to have again $p_{\text{tot}}=-c^2\rho$ at some subsequent period of time $\tau$, if this occurs, we will have $\bar{p}_{\text{tot}}=-c^2\bar{\rho}$ for any $u$, any $k$ and any $\chi$, as this is obtained from~\eqref{mat2}. This shows that such a compressed fluid will have the same thermodynamic properties in any reference frame.

Another important result is that spatial curvature $k>0$ eliminates the divergent character of the expressions of $(\bar{\rho },\bar{p})$ in the limit $u\to c$
\begin{equation}\label{mat3}
\bar{\rho }=-\frac{p_{\text{tot}}+c^2 (1-k \chi ^2) \rho }{c^2 k \chi ^2},\quad \bar{p}_{\text{tot}}=-\frac{c^2 \rho +(1-k\chi ^2) p_{\text{tot}}}{k \chi ^2}\,.
\end{equation}
Since $\bar{\rho }>0$ , we conclude from~\eqref{mat3} that, at those farthest galaxies where $u\to c$, the pressure $p_{\text{tot}}<-c^2 (1-k \chi ^2) \rho <0$ becomes negative (with $-c^2\rho<p_{\text{tot}}$). Since the thermodynamic pressure is $p=0$ at such distances (matter behaves as dust), this implies that $p_{\text{tot}}\simeq p_c<0$, and by~\eqref{p4} we have $\Gamma >0$,  which means that the production of matter and particles (what particles?) is still going on at such furthest galaxies as far as the Universe is expanding. This results in a repulsive gravity which may provide more impulse to the expansion, and may be causing cosmic acceleration, without invoking any of the dark components. In the language of $\Lambda$CDM, this corresponds to an effective cosmological constant $\Lambda_{\text{eff}}$, which in the flat FLRW model ($k=0$), drives the late accelerated cosmic expansion. To obtain the value of $\Lambda_{\text{eff}}$, we consider the extreme case with $u= c$ and $p_{\text{tot}}= -c^2 (1-k \chi ^2) \rho \simeq -c^2 \rho$, where we have dropped the term $k \chi ^2$. Substituting in~\eqref{eq2}, we obtain
\begin{equation}\label{mat3bb}
\frac{a''(\tau )}{a(\tau )}=\frac{8\pi G \rho(\tau)}{3}\,.
\end{equation}
In the flat FLRW Universe description, we can drop the matter content from the field equations in the limit $u\to c$ and obtain
\begin{equation}\label{mat3bbb}
\frac{a''(\tau )}{a(\tau )}=\frac{c^2\Lambda_{\text{eff}}}{3}\,.
\end{equation}
On comparing the two expressions~\eqref{mat3bb} and~\eqref{mat3bbb}, we obtain
\begin{equation}\label{ic2}
\Lambda_{\text{eff}}(\tau)=8\pi G\rho(\tau)/c^2\,,
\end{equation}
while at $\tau=0$ the expression was $\Lambda_{\text{eff}}=8\pi G\rho(0)/c^2$. Thus, in the limit of relativistic speeds and small (to negligible) pressures and densities, the SUM is equivalent to a flat FLRW model with an effective cosmological constant given by $8\pi G\rho(\tau)/c^2$ for $\tau \gtrsim\tau_0$ with $\tau_0$ being the present epoch (where in such limits, $\rho$ includes matter only). Similar to the flat FLRW model, the SUM justifies the late accelerated expansion too but with an effective cosmological constant that is function of time. But how does it change with time? It is worth noticing that the conclusion made here does not depend on the detailed evolution of $\rho$: All that we have considered are those farthest galaxies where $u\to c$ and where the thermodynamic pressure is 0. The conditions that have led to $\Gamma =3H_\Lambda$ in Sec.~\ref{secbounce} still hold, that is, we have $\Gamma(\tau) \simeq 3H(\tau)$ for $\tau \gtrsim\tau_0\,$.

To answer the last question we need to find how $\rho(\tau)$ changes with time. Equation~\eqref{p5} tells us that if the thermodynamic pressure is 0 (matter is dust at such distances), then
\begin{equation}\label{rho}
\rho'(\tau) = \Big(\Gamma(\tau) - 3~\frac{a'(\tau )}{a(\tau )}\Big)\rho(\tau) =[\Gamma(\tau)-3H(\tau)]\rho(\tau)\,.	
\end{equation}
We see that the sign of $\rho'(\tau)$ depends on the sign of $\Gamma(\tau)-3H(\tau)$. In this model a constant effective cosmological constant may be justified only if the rate of production is three times the Hubble parameter, that is, particle production compensates for three consecutive expansion rates. 

One may ask if the expansion may be halted. First of all $k$ is a very small number, it is unlikely that the r.h.s of~\eqref{p6m1} vanishes. However, if this happens at some time $\tau_\text{c}$, a (re)collapse process takes place. At $\tau_\text{c}$, we see from~\eqref{p6m1} that $\rho\propto a^{-2}$, which is similar to~\eqref{eq8} but different from it. Since $a'(\tau_\text{c})=0$, the derivative of the density vanishes also by~\eqref{p6} and $\Gamma$ vanishes by~\eqref{rho}, so the density becomes stationary about the instant $\tau=\tau_\text{c}$ (reaches its minimum value), from~\eqref{p6m2} we can determine the behavior of $a'(\tau)$ at $\tau=\tau_\text{c}\pm \Delta \tau$. $\rho$ can be approximated by $3c^2k/(8\pi Ga^2)$ and the thermodynamic pressure by 0 (this is justified as the Universe becomes more {\textgravedbl}dilute{\textacutedbl} at such scale factors). We obtain
\begin{align}\label{eq10}
&(a'(\tau))^2 = \frac{c^2k}{2}~\ln\Big(\frac{a_\text{c}}{a}\Big)\,,\quad [a<a_\text{c}=a(\tau_\text{c})]\,,\\
&H(\tau)^2 = \frac{c^2k}{2}~a^2\ln\Big(\frac{a_\text{c}}{a}\Big)\simeq \frac{c^2k}{2}~a_c(a_c -a)\,,\nonumber
\end{align}
which may be used to evaluate $k$ when the expansion reaches the (re)collapse point at $\tau_\text{c}$. Why in this case too does $\rho$ not follow the law $a^{-3}$? In SUM we assumed the spatial curvature $k$ to be a tiny small constant and could be neglected in most calculations. However, as $a\to a_\text{c}$, we can no longer neglect $k$, and this results in a violation of the Hubble law~\eqref{p6m1}, so that the expansion in this limit no longer implies $\rho\propto a^{-3}$.

\section{BHU versus SUM\label{secsvb}}
Both the BHU and SUM models of the Universe assume that the time evolution of the mass-energy configuration, making the whole content of the Universe, is confined into a finite space region, however, too large that the description of spacetime events within the cloud follows that of closed FLRW. In both models the formation of a singularity at the center of the configuration is halted by a negative degenerate pressure, $p=-c^2\rho$, resulting from the application of the Quantum Exclusion Principle to still undiscovered set of particles. This density is assumed to surpass that of the nuclear saturation density $\rho_{\text{nsd}}$. There is no known mechanism within particle-physics-current knowledge that gives a hint on when the bounce occurred under such high negative pressure (equivalent to an anti-gravity). This remark applies to Big Bang Theory as well: What caused the Big Bang and when did it occur? With that said, this situation resembles more the catastrophe theory where the moment of occurrence of some phenomena cannot be predicted (a famous example is the unpredictable timing and magnitude of a landslide). In the following we summarize the differences between the two models.

\begin{enumerate}
\item In the BHU the smallest radius of the cloud $r_b<r_\text{S}=2GM/c^2$ and in the SUM $r_b>r_\text{S}$, where $M$ is the total mass of the cloud. In the BHU the collapse is halted when all matter-energy content was confined in a space region where the radial coordinate is timelike. In the SUM, if collapse takes place, it is halted when all matter-energy content was confined in a space region where the radial coordinate is spacelike;
\item The BHU assumes that the mass-energy configuration remains regular for all $t$ and $r$ and yields the formation of a BH with \emph{one horizon} after collapse to a ground state. The SUM assumes that the mass-energy configuration remains regular for all $t$ and $r$ and yields the formation of no BH; of rather a star with no horizons;
\item In the BHU the region $r>r_\text{S}$ is no longer accessible to the mass-energy configuration once the ground state is formed. In the SUM the region $r>r_\text{b}$ beyond the mass-energy distribution is accessible;
\item In the BHU $r=r_\text{S}$ is a cosmological horizon, which is, as we have seen in Sec.~\ref{secbhu}, in contradiction with the fact that the bounce occurs in a region where $r$ is timelike ($r\leq r_b<r_\text{S}$). The bouncing fluid particles will move in the direction of increasing $r$ and, since $r$ is timelike, they will continue to move in that direction till they cross the surface $r=r_\text{S}$. The SUM yields a star with no horizons;
\item As to the effective cosmological constant driving the late accelerated expansion, the BHU yields a non-variable effective cosmological constant $3/r_{\text{S}}^2$ and the SUM yields a time variable effective cosmological constant $8\pi G\rho(\tau)/c^2$.  In SI units, taking $\rho (\text{now})=0.3\times 9.5\times 10^{-27}$, $c=299792458$, $G=6.673\times 10^{-11}$, $M_\odot =1.9888\times 10^{30}$, and $M=M_{\text{Univ}} =5\times 10^{22}M_\odot$ (mass of the Universe), we find $\Lambda_{\text{BHU}}=1.37588\times 10^{-52}$ and $\Lambda_{\text{eff}}(\text{now})=8\pi G\rho(\text{now})/c^2=5.3182\times 10^{-53}$. The SUM mimics standard dynamical dark energy models, which propose that the energy driving the universe's accelerated expansion changes over time. Recent measurements of baryon acoustic oscillation~\cite{ob2}, combined with with SNIa and CMB data~\cite{Abbott,ob1}, favor possible deviations from the standard $\Lambda$CDM, which hypothesizes the existence of a fixed cosmological constant with value $1.1\times 10^{-52}$ to $2.1\times 10^{-52}$ and offers no hint to derive it. This framework fixes the EoS parameter to $\omega=-1$, while dynamical models allow $\omega(z)$ to vary with the redshift $z(a)$~\cite{Jawad}, which is a function of the scale factor. In the SUM, the effective $\Lambda$ depends on $\rho(a)$,  which depends on the scale factor;
\item In BHU a positive spatial curvature $k$ was introduced by hand into the field equations. In SUM, since the initial metric at $\tau =0$ defined by~\eqref{s1dS}, \eqref{s3}, \eqref{s4} and $p_{\text{tot}}=-c^2\rho$ has already a constant positive spatial curvature, it was assumed that the evolution after the bounce occurred did not affect the value of $k$.
\item Both BHU and SUM rely on the existence of preons (and/or more fundamental particles). This conjecture implies that quarks and leptons are composite particles of more fundamental ones, which are still hypothetical. It is the degeneracy pressure of these more fundamental particles that halts the collapse to a singularity. Thus, the results derived in this work will remain true as far as the conjecture holds.
\end{enumerate}
\vskip10pt

\section{Conclusion -- How {\textgravedbl}closed{\textacutedbl} is closed?\label{secconc}}
We have emphasized that we cannot be living inside a BH. As an alternative to the BHU model, we assume that the matter constituents were tightly sewn together, so sewn that $p_{\text{tot}}=-c^2\rho$, in a ground state sourcing an SSRS of the field equations without a horizon. The ground sate is a star with radius larger than the star's Schwarzschild radius. The Universe was born when the matter constituents were torn apart, that is when the energy content of the de Sitter state starts getting converted to mass and particles. We do not believe that the hypothesis $p_{\text{tot}}=-c^2\rho$ is bold, as this is adopted in $\Lambda$CDM via the introduction of a scalar field. We depart from $\Lambda$CDM assuming that the Universe was not born from a singularity and that no associated inflaton field; rather, the Universe was born when its constituents, forming a regular star, were torn apart.

The birth was inflationary, a linear expansion occurred around the generalized Einstein static solutions, and the details of subsequent expansions should be qualitatively similar to $\Lambda$CDM counterparts.

We have shown that the density does not always drop proportionally to $a^{-3}$ and we provided two explanations for this behavior depending on the phase of expansion.

In our model, we assume the spatial curvature $k$ to be a tiny small positive constant not different from that of the initial de Sitter-static-metric state. Our argument is to assume that the bounce did not alter the value and sign of $k$. This allows us to depart from the standard flat assumption based on recent observations dwelling, with uncertainty, on positive $k$~\cite{ob1,ob2,ob3}. If $k$ is believed to be a small entity, it adds nothing to the news if one assumes it to be homogeneous (in a more realistic model): $k\equiv k(\tau)$. But how {\textgravedbl}small{\textacutedbl} is $k$ small; that is, how {\textgravedbl}closed{\textacutedbl} is the Universe closed? This is still an open topic for active research.

The proportionality formula of the effective cosmological constant and the late mass-energy density explains both the late accelerated inflation and the smallness of the effective cosmological constant.

\section*{Appendix: Matching conditions\label{secaa}}
\renewcommand{\theequation}{A.\arabic{equation}}
\setcounter{equation}{0}
We outline the procedure of matching the interior and exterior solutions for $\tau>0$ (\cite{junction,junction2,junction3} and references therein). For $\tau=0$, we already matched the interior de Sitter solution~\eqref{s1dS} with the exterior Schwarzschild line element and this yields the matching constraint~\eqref{s2}. The matching is possible because at $\tau=0$ the thermodynamic pressure of the interior solution is 0. As $\tau$ increases a boundary layer, described by a generalized Vaidya metric~\eqref{Vaidya}, develops to which the internal geometry is matched and which could in turn be matched to an exterior Schwarzschild geometry.

For $\tau>0$, the evolving internal metric~\eqref{sum5} can be brought to the form
\begin{equation}\label{a1}
\dd s_-^2=-c^2\dd \tau^2 + a^{2}(\tau)\dd \Psi^2 +a^{2}(\tau)\sin^2\Psi\dd\Omega^2\,,
\end{equation}
on performing the coordinate transformation $\chi=\sin\Psi/\sqrt{k}$ then replacing $a/\sqrt{k}$ by $a$. The evolving external metric is given by~\eqref{Vaidya} where $v$ is the retarded outgoing null coordinate and $r_v$ is the Vaidya radius. At the boundary $r_{\text{b}}$ we must have $a(\tau)\sin\Psi_{\text{b}}=r_v$ (equality of area radii). On the boundary surface $r=r_{\text{b}}$, the interior and exterior metrics reduce to
\begin{align}
&\dd s_-^2=-c^2\dd \tau^2 + a^{2}(\tau)\sin^2\Psi_{\text{b}}\dd\Omega^2\,,\\
&\dd s^2=-c^2\Big(\dot{v}^2-\frac{2GM(r_v,v)}{c^2r_v}~\dot{v}^2+\frac{2}{c}\dot{v}\dot{r_v}\Big)\dd \tau^2  + r_v^2\dd\Omega^2\,,
\end{align}
where $\dot{f}:=\dd f/\dd\tau$. Then, the first fundamental form constraint yields the following,
\begin{equation}\label{key}
\dot{v}^2-\frac{2GM(r_v,v)}{c^2r_v}~\dot{v}^2+\frac{2}{c}\dot{v}\dot{r_v}=1\,,
\end{equation}
with $r_v=a(\tau)\sin\Psi_{\text{b}}$.

Next, we need to match the second fundamental forms. For that end, the unit normal to the hypersurface $r=r_{\text{b}}$ are
\begin{widetext}
\begin{align}
	&n_-=(0,\,a(\tau),\,0,\,0)\,,\\
	&n_+=\Bigg(\frac{-1}{\sqrt{1-\frac{2GM(r_v,v)}{c^2r_v}+\frac{2}{c}\frac{\dd r_v}{\dd v}}},\,\frac{1-\frac{2GM(r_v,v)}{c^2r_v}+\frac{1}{c}\frac{\dd r_v}{\dd v}}{\sqrt{1-\frac{2GM(r_v,v)}{c^2r_v}+\frac{2}{c}\frac{\dd r_v}{\dd v}}}\,,0,\,0\Bigg)
\end{align}
\end{widetext}
The second fundamental form being defined by
\begin{equation}\label{key}
K_{ab}=\frac{1}{2}~(\partial_c g_{ab}n^c + g_{cb}\partial_a n^c + g_{ac}\partial_b n^c)\,,
\end{equation}
where ($a,\,b,\,c$) are spatial indices, we obtain from the continuity condition $\big[K^-_{\theta\theta }-K^+_{\theta\theta }\big]_{r=r_{\text{b}}}=0$
\begin{equation}\label{am1}
\cos\Psi_{\text{b}}=\frac{1-\frac{2GM(r_v,v)}{c^2r_v}+\frac{1}{c}\frac{\dd r_v}{\dd v}}{\sqrt{1-\frac{2GM(r_v,v)}{c^2r_v}+\frac{2}{c}\frac{\dd r_v}{\dd v}}}\,.
\end{equation}
The other continuity condition $\big[K^-_{tt}-K^+_{tt}\big]_{r=r_{\text{b}}}=0$ yields~\cite{junction2}
\begin{equation}\label{am2}
\partial_{r_v} M(r_v,v)=\frac{M(r_v,v)}{a(\tau)\sin\Psi_{\text{b}}}+a\ddot{a}\sin^2\Psi_{\text{b}}\,.
\end{equation}
The system of equations~\eqref{am1} and \eqref{am2} is not empty, as shown in~\cite{junction}, that is the system admits well defined solutions.



\end{document}